**Creativity in Science and the Link to Cited References:**

**Is the Creative Potential of Papers Reflected in their Cited References?**

Iman Tahamtan[1] & Lutz Bornmann[2]

1. Corresponding Author. School of Information Sciences, College of Communication and Information, University of Tennessee, Knoxville, TN, USA. Email: tahamtan@vols.utk.edu

2. Administrative Headquarters of the Max Planck Society, Division for Science and Innovation Studies, Hofgartenstr. 8, 80539 Munich, Germany. Email: bornmann@gv.mpg.de

Abstract

Several authors have proposed that a large number of unusual combinations of cited references in a paper point to its high creative potential (or novelty). However, it is still not clear whether the number of unusual combinations can really measure the creative potential of papers. The current study addresses this question on the basis of several case studies from the field of scientometrics. We identified some landmark papers in this field. Study subjects were the corresponding authors of these papers. We asked them where the ideas for the papers came from and which role the cited publications played. The results revealed that the creative ideas might not necessarily have been inspired by past publications. The literature seems to be important for the contextualization of the idea in the field of scientometrics. Instead, we found that creative ideas are the result of finding solutions to practical problems, result from discussions with colleagues, and profit from interdisciplinary exchange. The roots of the studied landmark papers are discussed in detail.

# 1 Introduction

Several scientometric studies have used cited references data for measuring novelty or creativity in science (e.g. Uzzi, Mukherjee, Stringer, & Jones, 2013; Wang, Veugelers, & Stephan, 2017). The general idea here is that if a paper contains many unusual combinations of cited references, the creative potential of the paper is high (Uzzi et al., 2013). Unusual combinations are those which can rarely be found in other publications. Uzzi et al. (2013) claim that combinations of atypical with conventional knowledge (reflected in unusual combinations) may lead to innovativeness (reflected in high impact papers). Therefore, creativity in science can be considered almost a universal phenomenon of conventionality and novelty (Uzzi et al., 2013). However, it is still not clear whether this approach of using cited references for measuring creativity is valid. The empirical studies based on patent data suggest that cited references might not have this prominent role (Callaert, Pellens, & Van Looy, 2014; Nagaoka & Yamauchi, 2015). We are not aware of any study which has similarly tested the approach for publications and cited references. The extent to which cited publications are the sources of inspiration leading to current landmark papers is not yet well understood. Thus, using exemplary landmark papers from the field of scientometrics, this study explores whether the creative potential of a paper is really reflected in its cited references or not.

# 2 Creativity in science

Creativity has been studied widely in the last decade. However, basic questions about the nature of creativity remain under debate (Kaufman & Beghetto, 2009). Numerous studies have attempted to analyze and understand what makes humans creative (Erren, Shaw, & Lewis, 2017). To study creativity and its nature, a group of studies and theories has investigated the lives of well-known and distinguished creators (e.g. Nobel laureates) by surveying or interviewing them – called *Big-C creativity* (Kaufman & Beghetto, 2009).

Intelligence and creativity are two vital ingredients in revolutionary science. Intelligence can be assessed by IQ tests. However, measuring the creativity of scientists is not an easy task (Charlton, 2009), despite the existence of psychological tests of creativity (Eysenck, 1995). The science of creativity aims to understand what leads to novel outcomes (Lee, Walsh, & Wang, 2015). Creativity, success in science, and scientific breakthroughs seem to be the result of several prerequisites such as interest among colleagues who take up on the ideas (Bornmann & Marx, 2012). Previous studies have used a variety of measures to operationalize the creative person in science, including self-reports, peer-ratings, scores on divergent thinking tests or personality inventories, total number of publications and citations, the h-index, etc. (Grosul & Feist, 2014). A variety of factors correlate with creativity, including affect, cognition, training, individual differences, culture, social behavior, team working, etc. (see Hennessey & Amabile, 2010). Individuals' characteristics such as intelligence, competency, motivation, knowledge, style, personality, etc. have been the core elements of creativity studies in a variety of fields (Shin & Jang, 2017).

A stream of research has examined the effect of team collaboration on the creativity of research, emphasizing that creativity is the outcome of individuals' interactions within a team (Shin & Jang, 2017). Interactions, collaborative networking, and information exchange might result in serendipitous discoveries in basic science, applied research and technological development activities (Yaqub, 2018). Uzzi et al. (2013) indicated that papers written in collaboration are 37.7% more likely than those of single authors to introduce novel combinations into conventional knowledge domains. Fleming (2007, p. 72) writes that "multidisciplinary collaboration increases the variance of the outcome, such that failures as well as breakthroughs are more likely". Lee et al. (2015) found that the probability of creative research is positively related to the increase in the number of members in a team, particularly when they have distinct knowledge domains. Lee et al. (2015) also point out the importance of the combination of diverse ideas for producing creative science.

Neumann (2007) interviewed 15 European Molecular Biology Laboratory (EMBL) group leaders (EMBL is a top non-US institution in terms of highly cited molecular biology and genetics publications). He found that scientists bring "together previously unlinked ideas to generate a new concept" (p. 203) or integrate previously puzzling fragments of information into a coherent picture. This study revealed that breakthroughs are largely internal, yet external factors also play an important role. Neumann (2007) notes that the crucial factors for stimulating creativity includes scientists' awareness of the unknown, colleagues' interactions, feedback after the emergence of a new idea, as well as environmental features. The social environment such as culture or team leader behavior can also influence individuals' motivation, and consequently creative performance (Hennessey & Amabile, 2010).

Erren et al. (2017) studied several creativity cases at Cambridge, and AT&T's Bell Laboratories. They found that working with 'an open door' leads to more thought exchanges, obtaining feedback, and enhancing individual and group creativity (Erren, 2008). Kasperson (1978) notes that colleagues are valuable sources of information, and creative scientists use such information sources effectively and differently from other scientists. Besides interaction with peers as a crucial factor for creativity (Neumann, 2007), the fostering of individual creativity is important too (Erren et al., 2017).

One important stream of research views creativity as involving the novel combination and/or recombination of elements that have never been combined before (Lee et al., 2015; Schumpeter, 1939). In this regard, Fleming (2001) maintains that the recombination of previously combined components and the combination of new components that have not been combined before could lead to creativity. "The combination process can be regarded as an innovation process that links knowledge with different distances in the knowledge base" (Zeng et al., 2017, p. 56). Atypical combinations of knowledge while maintaining the advantages of conventional knowledge could also lead to innovativeness (Uzzi et al., 2013). Carayol, Lahatte, and Llopis (2017) propose a measurement of novelty which uses the frequencies of pairwise combinations of articles' keywords for the exploration of new research questions. This study indicates that the combination does not always offer remarkable benefits for creativity, however pairwise keyword novelty is strongly related to the articles' citation impact (Carayol et al., 2017).

Koestler (1964) suggests the term 'bisociation' for human intellectual creativity. He explains this term as follows: "the creative act does not create something out of nothing, like the God of the Old Testament; it combines, reshuffles, and relates already existing but hitherto separate ideas, facts, frames of perception, associative contexts" (Koestler, 1981, p. 2). A similar idea is introduced by Salganik (2017) called 'ready-made'. Ready-made characterizes art "where an artist sees something that already exists in the world and then creatively repurposes it for art" (Salganik, 2017, p. 7). Salganik (2017) also introduces another term called 'custom-made' which is defined as the art that was intentionally created.

Schubert (2013) employed bisociation and proposes 'title term bisociation' in bibliometrics as a tool for detecting emergent areas. Schubert and Schubert (1997) posit that emergent areas or new connections occur if "two frequent but so far not co-occurring terms begin to co-occur regularly" (p. 132). Schubert and Schubert (1997) analyzed the terms in the titles of documents in *Inorganic Chemistry Acta* and found 14 new connections. They note that "using Koestler's 'bisociation' concept, some potential 'creative foci' were identified in the form of pairs of title terms, around which some new ideas may emerge" (Schubert & Schubert, 1997, p. 133).

Creative ideas and discoveries are not always related to the above-mentioned factors. Sometimes researchers make unexpected and beneficial discoveries, an occurrence termed serendipity (Yaqub, 2018). Serendipity takes different forms. For example, researchers working on one problem may make their discovery in another. Sometimes research finds a solution for a given problem via an unexpected route. In other cases, investigations lead to the solution of an entirely different problem (Yaqub, 2018).

# 3   Overview of concepts for using cited references to measure creativity

Several approaches have been proposed in recent years to measure the creativity (or novelty) of papers on the basis of (pairs of) cited references (e.g. Uzzi et al., 2013; Wang et al., 2017). Such an approach for measuring creativity could be supported by the suggestion that many new ideas in science are inspired by previous studies (Zeng et al., 2017), or that new combinations of previous components lead to creative ideas (see section 2). The cited references in patent documents have also been used for tracing the scientific sources of innovations (Nagaoka & Yamauchi, 2015). Although several studies have used (pairs of) cited references for creativity measurements, the validity of this approach has not really been tested: is the creativity of ideas which appeared in publications really reflected in their cited references? To the best of our knowledge, this question has not been answered yet.

Uzzi et al. (2013) investigated the extent to which scientific papers cite novel versus conventional combinations of prior work. They divided the references of papers into conventional and novel combinations. They counted the frequency of co-citations of journal pairs in the bibliography of nearly 18 million research articles from the Web of Science (WoS, Clarivate Analytics) to determine whether any particular pair of cited references is conventional or atypical. Journal pairs were considered novel if they were cited together for the first time in the literature. They calculated the actual and expected frequency for each co-cited journal pair. The observed frequency of journal pairs was then compared to the expected frequency resulting from randomized citation networks. Actual and expected frequency counts for each journal pair were converted into a z-score.

In the study, "z scores above zero indicate pairs that appeared more often in the observed data than expected by chance, indicating relatively common or 'conventional' pairings. z scores below zero indicate pairs that appear less often in the observed WOS than expected by chance, indicating relatively atypical or 'novel' pairings" (Uzzi et al., 2013, p. 469). They then calculated 10th percentile z scores (left tail) and median z scores. The median z score for an article was used to characterize the paper's conventionality, and the left tail to characterize the paper's more unusual combinations, i.e. papers with "high novelty". Results indicated that more than half the papers had median z scores above 64, and 41% of the papers had a 10th-percentile z score less than 0. This study found that the highest-impact papers primarily relied on conventional combinations of prior work, and rarely incorporate atypical combinations. They noted that "papers with an injection of novelty into an otherwise exceptionally familiar mass of prior work are unusually likely to have high impact" (Uzzi et al., 2013, p. 470). This study also found that papers with a combination of novelty and conventionality are twice as likely to be highly cited works.

Boyack and Klavans (2014) replicated the study of Uzzi et al. (2013) by using a slightly different methodology. They used 12 million articles and conference papers in Scopus from 2001-2010 rather than investigating WoS research articles. In addition, rather than using z scores for calculating novelty (atypical combinations) and conventionality (typical combinations) (see Uzzi et al., 2013), they "calculated K50 statistics for co-cited journal pairs. The difference is that the expected and normalization values for K50 are calculated using the row and column sums from the square co-citation count matrix rather than using a Monte Carlo technique" (Boyack & Klavans, 2014, p. 65). Articles with K50 values below 0 were considered atypical combinations. They maintained that the method of using journal pairs for measuring novelty might have some potential problems, because it does not consider differences across fields (disciplinary effects) and journals (journal effects). For example, the three multidisciplinary journals of *Nature*, *Science*, *Proceedings of the National Academy of Sciences of the United States of America* (PNAS) accounted for 9.4% of all atypical combinations, suggesting the possibility of the existence of significant journal-level effects. However, Boyack and Klavans (2014) believe that the methodology used by Uzzi et al. (2013) for measuring creativity in science is sound and such issues do not invalidate it. Nevertheless, they maintain that other novelty measures that are relatively independent of discipline and journal effects should be introduced (Boyack & Klavans, 2014).

Wang et al. (2017, p. 1417) defined *scientific novelty* "as the recombination of pre-existing knowledge components in an unprecedented fashion". They analyzed 661,643 unique research articles published in 2001 across all scientific disciplines in the WoS (Veugelers & Wang, 2016; Wang et al., 2017). They classified novelty into three categories: (1) non-novel: papers without new journal combinations; (2) moderately novel: papers with at least one new combination and with a novelty score lower than the top 1% of their subject category; and (3) highly novel: papers that make more distant new combinations; papers with a novelty score among the top 1% of their subject category (Veugelers & Wang, 2016; Wang et al., 2017). They followed Uzzi et al. (2013) for measuring novelty in science by taking into account journal pairs that were cited in publications. However, rather than looking at the atypicality of referenced journal pairs, they measured novelty by examining the number of new journal pairs in the references of a paper weighted by the cosine similarity between the newly-paired journals (Veugelers & Wang, 2016; Wang et al., 2017). They introduced a difficulty score to capture the difficulty of combining two journals. The 'difficulty' of making the new combination was measured by taking into account the knowledge distance between the newly-combined journals based on their co-cited journal profiles.

Wang et al. (2017) found that only 11% of papers make at least one new journal combination in their reference lists, among which 54% make only one, and 7% contain more than 5 new combinations. Their results also indicated that "most of the novel papers score only modestly on [their] distance-weighted novelty indicator" (p. 1419). Highly novel papers are those that make more distinct new combinations. Such papers are less cited when using a short citation time window, but are more likely to place among the top 1% highly cited papers, when using a sufficiently long-time window (Wang et al., 2017). Moreover, novel papers are more likely to receive a large number of citations in other than their own fields.

Patents are frequently used in scientometrics to measure innovations. Veugelers and Wang (2016) identified which type of science (novel or non-novel) is most likely to be cited by patents. They measured 'scientific novelty' based on the number of new combinations of referenced journals. The links between science and technology were assessed by measuring the number and type of non-patent-references in patents (the references in patents which were journal articles). The patent dataset used in this study was all the patents in PATSTAT – a database containing patent data from leading industrialized and developing countries. Using an algorithm developed by the Centre for Research & Development Monitoring (ECOOM), they matched non-patent-references in patents to individual publications in WoS. The results showed that only 10% of all publications were referenced in patents (technological inventions), "but a small number of scientific papers which score on novelty (about 11%) are significantly more likely to have technological impact, particularly the 1% highly novel scientific papers" (Veugelers & Wang, 2016, p. 270).

Callaert et al. (2014) interviewed 33 Belgian inventors to investigate the sources of inspiration and the extent to which the non-patent references in patents were the sources of inspiration leading to the current inventions. Their results indicated that scientific references (non-patent references) act as a source of inspiration for about 50% of the inventions. However, 30% of the patents that were inspired by scientific references didn't contain any scientific reference. Furthermore, half the scientific references in patents were evaluated as unimportant or background information by the inventor. The authors conclude that scientific references might be related to the citing patent, however the patent might not necessarily have been inspired by these references (Callaert et al., 2014). We found a similar result in the current study. The reasons why scientific documents might be read but not necessarily cited are explained in Tahamtan and Bornmann (2018).

Nagaoka and Yamauchi (2015) asked Japanese inventors whether and which scientific sources were essential and important for the conception or implementation of their invention. Their results indicated that half the patents relied on scientific references in their research processes, among which only 17% revealed such important literature in their patent documents. This study also indicated that in the inventions that cite scientific literature in their patent documents, these references were not essential for the conception or implementation of 82% of the inventions. Therefore, scientific references were essential for the conception and implementation of R&D in only 18% of inventions. According to this study, cited references are not good measures for tracing the scientific sources of innovations (Nagaoka & Yamauchi, 2015). It seems that cited scientific references are not necessarily essential for the patented invention, but might serve as background information and a source of inspiration for the invention (Veugelers & Wang, 2016).

While some previous studies show the existence of a positive relationship between certain combinations of cited references and the impact of the citing papers, the link to innovativeness and creativity is not clear. It is not well understood whether the creativity of papers is rooted in their cited references.

## 4   Research objectives

To investigate whether the creative potential of scientific papers is reflected in their cited references, we surveyed several authors who have published landmark papers including important concepts in the field of scientometrics. We asked how the idea for the landmark paper is reflected in their selected cited references. Since the authors did not only focus in the answers on their use of cited references, we also received information on the context in which the landmark papers emerged. The authors explained in detail how the idea originated and what the conditions for the idea were. They described in which concrete situation the idea arose. For example, the idea of using the number or proportion of papers belonging to the 10% most frequently cited for research evaluation purposes was already born at the beginning of the 1980s. Francis Narin used this indicator in his doctoral thesis and in research evaluation studies at the National Institutes of Health (NIH). This means that the currently preferred field-normalized indicator in bibliometrics (Hicks, Wouters, Waltman, De Rijcke, & Rafols, 2015) was probably the first introduced; field-normalized indicators based on mean citation rates – the standard over decades – have been proposed later in the mid-1980s by the Budapest group of bibliometricians.

## 5   Methods

This is a qualitative empirical study in which we investigated the origin(s) of creative ideas in scientometrics with a focus on cited references. In the first step, we identified very important concepts in scientometrics and the corresponding landmark papers. The concepts were not selected systematically, based on a structured search of the literature. Instead, a few concepts were identified based on the long experience of one of us in scientometrics and the reading of various overviews of the literature in scientometrics. From this unsystematic selection approach it follows that the results of this study cannot be generalized. They can only point to possible problems with the approach of measuring creativity based on cited references. Larger studies based on more systematically composed sets of publications should follow.

The concepts and papers which we included in this study are as follows:

We asked the corresponding author of each paper about the origin(s) of the creative idea, how the author came to the idea, and how this idea is reflected in the cited references of the paper. We asked each corresponding author how the creative process is reflected in their choice of cited references. We further asked – if the process was not (completely) reflected in cited references – what the possible reasons for coming up with the creative idea were.

The free texts (responses) were analyzed by both authors of this paper. Overall themes were distinguished, and then interpreted in the context of the origin(s) of creative ideas. Throughout the empirical analyses, we constantly discussed the interpretation of the responses by the authors. Disagreements were resolved by negotiation and consultation.

The email text which we sent to the authors of the landmark papers was as follows:

"Dear [corresponding author],

In recent years, it has been proposed to measure the creative potential of papers based on their cited references. I guess you know, for example, the paper by Uzzi and co-authors published in *Science* ("Atypical combinations and scientific impact"). The basic idea is as follows: if a paper contains many unusual combinations of cited references, its creative potential is high. Thus, the proposal for measuring creativity is based on the premise that the cited references reflect creativity in any form. However, it is still not clear whether this approach can really measure creative potential of papers or not. Thus, we have started a study to understand in more detail whether the creative potential of a paper is reflected in its cited references.

In the first step of our study, we identified landmark papers which introduced key concepts in scientometrics. Your study, [corresponding author], introducing [key concept] in the field is of course a landmark paper. According to our search of the literature, you introduced this method in the following paper: [bibliographic information of the paper]. Please correct us, if this is not true.

We are interested in the origin of your idea for the method, how you came to the idea and how this idea is reflected in the cited references in your paper. Below you can find the cited references which we have extracted from the paper:

[List of references of the paper]

Could you please describe in a (short) free text how the idea for the [concept] emerged? Furthermore, we are interested in how this process is reflected in your choice of cited references. If the process is not reflected in cited references, we are interested in possible reasons for that (for example, because you developed the idea in personal communications).

If you have questions about our study, feel free to contact us [our names and contact data]. If you have any problems with using your free text for the upcoming paper on our study, please let us now. It is clear that we abstain from using your landmark paper then".

## 6   Results

In this section, we discuss the breakthrough concepts in scientometrics and the corresponding participants' responses to our email. We discuss what the innovative potential of these concepts is, how they have been applied in practice, which problem(s) could be solved by the concepts, and their potential drawbacks, if any. Many concepts refer to field-normalization of citation impact which is the standard in bibliometrics for cross-field comparisons. Cross-field comparisons of citation counts are profoundly influenced by the differences in citation behaviors of scientific fields (Schubert & Braun, 1986). For example, mathematics has a much lower citation density than the biomedical sciences (Leydesdorff & Opthof, 2010). Therefore, citation counts are inadequate scientific measures for evaluative cross-field comparisons, unless we remove the field effects (Waltman, 2016).

In comparisons among different fields, field normalization of citation impact has been widely employed in bibliometric studies. One challenge in the normalization process has been constructing proper scaling procedures (Schubert & Braun, 1996). A possible solution could be using percentiles for field normalization. Another solution is fractional counting of citation impact in the citing articles.

In the following, we will explain the breakthrough concepts in scientometrics, including Relative Citation Rate (RCR), the Characteristic Scores and Scales (CSS) method, top-decile performance (number or proportion of papers belonging to the 10% most frequently cited papers in a field), CPP/JCSm and crown indicator (CPP/FCSm) – based on journal and field normalization, citing-side normalization (audience factor), co-citation analysis, h-index, and altmetrics.

## 6.1 Relative Citation Rate (RCR)

### 6.1.1 Explanation of the concept

Before the introduction of the RCR, the mean citation rate had been frequently used as an indicator for measuring the citation impact of a given set of publications (Schubert, Glänzel, & Braun, 1983). However, the mean citation rate varies greatly between fields; thus it is not a proper indicator for comparing publications from different fields (Schubert et al., 1983). Therefore, to make cross-field comparisons of scientists, groups, institutions, or countries possible, relative indicators have been proposed (Schubert & Braun, 1986). Schubert et al. (1983) were among the first scholars who proposed a relative indicator called RCR as a field-normalized indicator.

RCR relates the observed citation rate of a set of publications to the expected (Schubert et al., 1983). It is defined as the ratio of the observed citation rate of a set of papers to the expected citation rates of the journals where the papers were published. In other words, RCR "measures the citation impact of a given set of publications (e.g. the publication output of countries or research institutions) as related to the respective world average" (Schubert et al., 1983, p. 80).

RCR = 1 indicates that the set of publications are cited exactly at an average rate; RCR > 1 indicates that the citation rate of the publications is above the expected average (Schubert, Glänzel, & Braun, 1989).

### 6.1.2 Qualitative analysis based on the email in Appendix 8.1

András Schubert came up with the idea of RCR long before he became a "professional" in scientometrics when he was thinking about the possibility of cross-field comparisons by citation indicators. The idea emerged in a situation when he was not working on a particular bibliometric problem. The discussions with his colleague, Imre Ruff, a fellow professor of chemistry, were an important element in developing the idea. Putting the idea into practice seems to have been quite a laborious task because at that time only the printed volumes of the Science Citation Index were available. However, many hours spent tallying citations led to the first results. RCR was first used in practice in the Hungarian Academy of Sciences for comparing the relative standing of Hungarian institutions. These results were not published as they were considered confidential.

The first publication of "relative impact" appeared in 1982 in Hungarian and the authorship was not entirely clear. An English publication followed three years after the first publication of the idea. The statistical knowledge of Wolfgang Glänzel led to the addition of a statistical reliability estimation to the indicator. The term "Relative Citation Rate" was first used for the indicator as a new bibliometric tool in the 1983 Varna conference. The idea eventually gained its final form and place in a paper published in *Scientometrics* in 1986. In the paper presented on the Varna conference, no reference was listed due to the strict space limitations. Independent of this formal requirement, it seems that the idea of RCR was not the result of reading and using (bibliometric) literature.

## 6.2 Characteristic Scores and Scales (CSS) method

### 6.2.1 Explanation of the concept

The method of characteristic scores and scales (CSS) was introduced by Schubert, Glänzel, and Braun (1987a) for assessing research performance, and identifying papers with excellent and outstanding citation rates. CSS can also be used to rank journals in terms of the amount of highly cited papers they publish, and to identify top journals as well as highly cited papers within scientific fields (Glänzel, 2011). Glänzel (2011) maintains that CSS can be used to compare the citation patterns of subsystems directly with citation patterns of the complete system. He writes that "the method of CSS can be applied to different aspects of the same system such as to classify individual publications or journals within the same subject even if their citation distributions follow different models" (Glänzel, 2011, p. 48). The CSS method can not only be used for the classification of single papers, but also the classification of institutions in meaningful groups, and of highly-cited researchers (Bornmann & Glänzel, 2017). The CSS method has been also applied to citations to compare how altmetric events (Mendeley readership counts, tweets and blog mentions) differ from citing patterns (Costas, Haustein, Zahedi, & Larivière, 2016).

Vîiu (2017, p. 749) writes that the cornerstone of the CSS method "is that of allowing a parameter-free characterization of citation distributions in such a way that impact classes are defined recursively by appealing to successive (arithmetic) means found within a given empirical distribution". Therefore, the approach used in CSS addresses one of the fundamental problems in bibliometrics – the skewness of science (Vîiu, 2017). Another advantage of CSS is that "it is self-adjusting, that is, no predefined thresholds, such as given numbers of citation or percentiles, are necessary" (Glänzel, Debackere, & Thijs, 2016, p. 2).

The application of the CSS method is as follows: each paper in a database should be classified into one field. The following characteristic scores (thresholds) of the citation distributions within one field are determined (Schubert et al., 1987a):

- XO is identically zero

- X1 is the mean citation rate per paper

- X2 is the mean citation rate of papers cited above average

- X3 is the mean citation rate of papers cited above X2

The characteristic scores are used to group the papers into five categories of citedness (Schubert et al., 1987a), and to distinguish between poorly, fairly and highly cited papers (Glänzel, 2011). The five categories of citedness are as follows:

- Category 0: not cited papers

- Category 1: poorly cited papers: papers cited lower than average (XO, X1)

- Category 2: fairly cited papers: papers cited at least average but below X2

- Category 3: remarkably cited papers: papers cited not lower than X2 but below X3

- Category 4: outstandingly cited papers: papers cited not lower than X3

Based on the above classification scores, two diagrams are drawn for each field. The first diagram is a linear scale featuring the characteristic scores, XO, XI, X2, X3; and the second one is a horizontal bar chart indicating the percentage distribution of papers between the categories of citedness 0 through 4 (see Schubert et al., 1987a). Using this method, the papers in a field can be classified as "not cited" (this class is no longer considered in the CSS method) "poorly cited", "fairly cited", "remarkably cited" and "outstandingly cited" (Bornmann & Glänzel, 2017). The last two groups refer to successful research activities in terms of citations. In addition, "the share of papers published by units (e.g. researchers or institutions) in these impact categories can be determined" (Bornmann & Glänzel, 2017, p. 1078).

### 6.2.2   Qualitative analysis based on the email in Appendix 8.2

Wolfgang Glänzel developed the CSS idea as a professional bibliometrician. He noted that the idea of CSS is based on several influences which are mainly rooted in the mathematical foundations of bibliometrics. One element was the analysis of citation distributions of journals. He found that while the mean citation rates of journals covered in the Science Citation Index database are similar, their shares of cited papers (or their not cited papers) are sometimes quite different. This was closely connected with the second basis of CSS: his personal research interest in probability theory that was focused on the analysis of probability distributions. A further important basis were the main findings in his dissertation in mathematics related to the modeling of stochastic processes in bibliometrics in 1984. He believes that these findings could be considered the main theoretical basis without which the CSS method would probably never have been developed in its present form.

He couldn't remember how the idea of truncating the distributions at their mean value and of repeating this procedure on the iteratively truncated samples emerged. Because of the relationship with the characterization theorem published in Glänzel et al. (1984), the method was called "Characteristic Scores and Scales" [this paper is not cited in the corresponding papers introducing the method]. The CSS method was first used for journal analysis. The very first version of the method used the auxiliary class of not cited papers, but the class was later dropped. The idea of the CSS method was initially published twice. The method was more or less neglected over many years, probably because of the enormous computational demands. The robustness and advantages of the method grew twenty years later, when hard- and software development and database access allowed a broadening of the scope beyond journal analysis and the application of the method to larger citation windows (20 years or even more).

## 6.3 Top-decile performance (number and proportion of papers belonging to the 10% most frequently cited papers in a field)

### 6.3.1 Explanation of the concept

Top decile performance, proposed by Narin (1987, p. 102), is defined as "the percent of an institute's papers which are amongst the more highly cited 10% of papers in a field". This method uses a field-dependent threshold (i.e., the 10% most highly cited publications) to determine whether a paper is highly cited or not (Waltman, 2016). The indicator has been proposed to measure the percentage of papers in the top decile of a subfield or even a group of combined subfields (Narin, 1987). The advantage of the top decile performance is that the differences in field-specific citation practices scarcely influence the measure – similarly to the RCR and CSS methods. Hicks et al. (2015, p. 430) note that "the most robust normalization method is based on percentiles: each paper is weighted on the basis of the percentile to which it belongs in the citation distribution of its field (the top 1%, 10% or 20%, for example)".

The idea of calculating the proportion of publications that belong to the 10% most highly cited in their field is used in both the CWTS Leiden Ranking (here it is called $PP_{top10\%}$) and the SCImago Institutions Rankings (Waltman, 2016). In 2011, the third edition of the SCImago Institutions Rankings (SIR) World Reports introduced the so called Excellence Rate, which "provides the percentage of papers published by an institution belonging to the top-10% papers in terms of normalized for the same field of publications and the same publication year" (Bornmann, de Moya-Anegón, & Leydesdorff, 2012, p. 333). One critique of the top-10% indicator is its arbitrary threshold for selecting highly cited papers; for instance, why not use the top 5% or top 20% most frequently cited publications (Waltman et al., 2012)? Tijssen, Visser, and Van Leeuwen (2002) used the top 1% and top 10% most highly cited publications, and Van Leeuwen, Visser, Moed, Nederhof, and Van Raan (2003) the top 5% most highly cited publications.

Another problem of the top-10% approach is ties of citation numbers around the threshold of 10%. Due to ties, it is difficult to assign some publications in a field to the top 10% or not (Waltman et al., 2012). Suppose 100 papers: 5 papers having 100 citations each, 10 papers having 80 citations each and the rest with 20 citations each. It is not clear whether the 10 papers with 80 citations should belong to the 10% most-cited papers or not. One solution is proposed by Waltman et al. (2012), assigning these papers fractionally to the number of top-10% papers.

### 6.3.2   Qualitative analysis based on the email in Appendix 8.3

The idea of the top-decile performance indicator was based almost entirely on the doctoral dissertation of Francis Narin, which was written with the cooperation of the National Institute of Health (NIH) and with an NIH based data set. The dissertation entitled "Concordance between Subjective and Bibliometric Indicators of the Nature and Quality of Performed Biomedical Research" (Narin, 1981) contained an extensive discussion of bibliometric data and its significance in the NIH evaluations context. One basis of the idea for the indicator was the problem they encountered in looking at the citation impact of the papers at NIH: the citation densities of subject areas within biomedicine varied radically. For example, the citation density in dental research was far lower than in biochemistry. To resolve this issue, they had to devise an indicator of research excellence that was independent of the citation density of different biomedical areas. The way they thought of doing this was by looking at the top 10% of most highly cited papers, the top decile, and comparing the different NIH institutes in terms of their top decile performance in these specific areas. Francis Narin wrote that they just invented things when they needed them in evaluative practice. Thus, the idea for the top decile indicator did not derive from the literature and is not reflected in the cited references.

### 6.4 Crown indicator (CPP/FCSm)

#### 6.4.1 Explanation of the concept

The Center for Science & Technology Studies (CWTS) at Leiden University introduced the crown indicator (CPP/FCSm) for field normalization (Moed, De Bruin, & Van Leeuwen, 1995a). The indicator was intended to set the standards for the field of bibliometrics and has been widely used in research evaluations (Leydesdorff & Opthof, 2011; Opthof & Leydesdorff, 2010). The indicator is similar to the RCR (see section 6.1). Values below (above) one indicate that the performance of a set of papers is below (above) the international standard of the corresponding field (Opthof & Leydesdorff, 2010).

In CPP/FCSm, CPP (citations per publication) is the average number of citations per publication for a set of papers (from a research group or institution), and FCSm or (sub)field citation score is the average number of citations of the subfields (journal categories) in which the papers have been published (Moed et al., 1995a). CPP/JCSm is a similar indicator developed by CWTS based on journal normalization. JCS (Journal Citation Score) is the "average number of citations per publication for a particular type of article published in a particular journal in a specific year" (Moed et al., 1995a, p. 421). JCSm, the mean journal citation score, is the average number of citations of the journals in which the evaluated set of papers has been published. What makes CCP/JCSm distinct from RCR (see Schubert et al., 1983) is that CPP/JCSm takes into account both the document type and the year in which the paper has been published. Therefore, CPP/JCSm can be considered a more accurate indicator than RCR for normalized citation impact (Moed et al., 1995a). However, Moed et al. (1995a) also suggest that FCSm is the most appropriate expected value to be used for field-normalization, since single journals do not represent fields.

Despite the extensive use of the crown indicator for bibliometric evaluations (see Bornmann, Mutz, Neuhaus, & Daniel, 2008), its flaws have been discussed in previous studies (Bornmann

& Mutz, 2011; Leydesdorff & Opthof, 2010; Lundberg, 2007). For example, Opthof and Leydesdorff (2010, p. 424) maintain that "the field normalization [CCP/FCSm] is more problematic than the journal normalization [CCP/JCSm] because the subject categories of the ISI [today Clarivate Analytics] sometimes heavily overlap and are often misguided." One suggestion to overcome this issue is the use of field classifications of the Library of Congress (Bensman & Leydesdorff, 2009) or Medical Subject Headings (MeSH) of the bibliographic database MEDLINE (US National Library of Medicine) (Bornmann et al., 2008). Another criticism mentioned by Opthof and Leydesdorff (2010) is related to the order of operations in CCP/JCSm and CCP/FCSm in which 'divisions' precede 'additions'. They propose that instead of comparing citation rates with the average citation rates of journals (JCSm) and fields (FCSm), averaging the normalizations on a paper-by-paper basis should be used (Opthof & Leydesdorff, 2010). Similarly, Lundberg (2007) notes that the crown indicator should normalize on the level of individual publications.

### 6.4.2 Qualitative analysis based on the email in Appendix 8.4

Henk Moed writes that the idea of comparing the citation impact of a unit of assessment (UoA, e.g., author or group) with that of a particular benchmark set, and calculating the ratio between the two citation rates, was not new. The ISSRU group in Budapest did that already in the early 1980s and named their indicator RCR (see above). The Leiden report had also done so in Moed, Burger, Frankfort, and Van Raan (1983). The RCR paper and the report are not cited in Henk Moed's paper. Henk Moed noted that there were various ways to define the benchmark set. For example, ISSRU compared the average citation rate of a country's articles with that of all papers published in the journals in which that country had published, which is a variant of the CPP/JCSm. Vinkler (1986) defined a relative citation rate using journal categories as benchmark sets. Henk Moed and colleagues generalized the various versions of the RCR, introducing a citation balance model in the early 1990s. CPP/FCSm was created based on the premise that single journals cannot represent complete fields.

Henk Moed writes that although their field-normalization method is based on previous research, the citation balance model was original. What is probably also original is that the previous RCR did not take into account (correct for differences in) the subject field (journal category), the year of publication, and the document type.

Henk Moed and colleagues have cited some of the resources mentioned in their 1995 landmark paper (Moed, De Bruin, & Van Leeuwen, 1995b). Thus, the crown indicator is based on previous research (which is mostly cited), but they have cited also a few "setting the stage" papers.

## 6.5 Citing-side normalization (audience factor)

### 6.5.1 Explanation of the concept

One of the most recently introduced normalization approaches of citation impact is citing-side normalization. Zitt and Small (2008) proposed the concept for the purpose of reducing the field dependence of the journal impact factor (JIF). The audience factor (AF) "is a variant of the journal impact factor where emitted citations are weighted inversely to the propensity to cite of the source" (Zitt, 2010, p. 392). In other words: "the basic definition of the audience measure AF relies on a weighting of emitted citations inversely to the average length of bibliographies in the source, either strictly at the journal level (citations from the journal are weighted in inverse proportion of the bibliography length in the journal) or at the field level (citations from the journal are weighted in inverse proportion of the bibliography length in the field where the journal belongs)" (Zitt, 2010, p. 393). Citing-side normalization is based on the idea that the citation density differences in fields are influenced by the differences in the length of publications' reference lists. Thus, the field normalization approach aims to normalize citation impact by correcting for the effect of the length of publication's reference lists (Waltman, 2016).

Unlike cited-side normalization approaches, which are focused on citations of the cited paper, the audience factor is focused on the citing journals (Zitt & Small, 2008). The approach "takes into consideration the citing propensity of journals for a given cited journal, specifically, the mean number of references of each citing journal, and fractionally weights the citations from those citing journals" (Zitt & Small, 2008, p 1856). The audience factor is a weighted form of the JIF, which normalizes the citing propensity differences, and immediacy among fields (Zitt & Small, 2008). Numerous studies have adopted the concept of citing-side normalization to create various citing-side field-normalized indicators similar to the audience factor. These indicators are explained, for example, in Waltman and van Eck (2013). The empirical study by Waltman and van Eck (2013) demonstrates that these indicators outperform traditional cited-side indicators.

### 6.5.2    Qualitative analysis based on the email in Appendix 8.5

Michel Zitt distinguishes the various tracks leading to Zitt and Small (2008). He notes that references cited in the bibliography of the paper are in bold face in his explanatory text. He notes that the 2008 article did not introduce the idea of citing-side normalization (fractionation). This idea already existed in three distinct research lines in the 1970s. However, the aim of their 2008 paper was to characterize the features of the method as a general tool of citation normalization with a focus on journal impact.

He notes that the references of their 2008 article are important for the background of citing-side normalization/fractionated citations, among which Pinski and Narin (1976) was the most important one (influence weights are proposed in this paper). He appears to state that they had to sacrifice some references due to the *Journal of the American Society for Information Science*' rules on bibliography length for non-original articles (their 2008 article was not published as a full article).

Michel Zitt met Henry Small at two conferences, Leiden STI 2006 and Madrid ISSI 2007, where they discussed the citing-side idea, which resulted in the publication of the 2008 article. The main sources of audience factor are as follows:

1. Forerunners

The influence of the bibliography length on citation impact had been recognized quite early in Philadelphia. Michel Zitt refers to one of the references in their paper: Small and Sweeney (1985), who applied fractional counting in co-citation indices, refer to two independent forerunners: Tyler Thompson of Rutgers University and Martha Dean of ISI, who proposed fractionation of citations in 1976 to overcome the problem of bibliography length bias.

2. Influence weight

The same year, Francis Narin and colleagues at Computer Horizons Inc. (CHI) published a book (Narin, 1976) which stressed the problem of referencing, and another classic paper which introduced influence weights (Pinski & Narin, 1976), one of the references in their 2008 paper. These resources, specifically Pinski and Narin (1976), were important sources for the idea of the audience factor.

3. Works on diversity and critique of conventional normalization methods

Michel Zitt and Henry Small started from an instantiation of Pinski and Narin (1976) on the ground that it was not necessary to reinforce the Matthew effect implied by iteration of sources´ prestige. Therefore they just retained from Pinski and Narin (1976) the built-in referencing correction and not the recursion. They focused on efficient normalization methods in practice in the context of the issue of field diversity. The recognition of the diversity of citing behavior was the motivation of forerunners quoted above, such as Francis Narin's team, and Murugesan and Moravcsik (1978). Also, Garfield (e.g. 1979, in the reference list of their 2008 paper) repeatedly stressed the referencing bias, rather than for example the size of fields, as a determining factor of the impact level. Michel Zitt notes that they cited Biglu (2008) in their paper just as testimony of the continuous flow of works in this line.

Another basis for the audience factor was Zitt and Bassecoulard (1994, 1996) in which a kind of fractional measure was used in their co-citation studies [1996 is not cited in their 2008 paper].

The limitations of classical indicators of citation impact was another basis for developing the audience factor: the field dependency of impact and the necessity of field normalization. Several works of Michel Zitt and colleagues addressed these limitations. For example, Zitt, Ramanana-Rahary, and Bassecoulard (2005) [cited in the 2008 paper] addressed the instability issue of the classical cited-side normalization methods.

## 6.1    Co-citation analysis

### 6.1.1    Explanation of the concept

The concept of co-citation has been introduced independently by Small (1973) in the U.S. and Marshakova (1973) in Russia. Co-citation is defined as the frequency with which two documents are cited together by a later document. In this concept, "the number of identical citing items defines the strength of co-citation between the two cited papers" (Small, 1973, p. 265). In other words, the strength of co-citation is determined by the number of documents that have cited two earlier works. Therefore, co-citation depends on the citing authors in the sense that the more citing authors two earlier works jointly cite, the stronger the relationship of the two earlier works (Small, 1973). Co-citation patterns represent the relationship between cited documents, as well as the ideas, concepts, methods, etc. presented in these documents. Co-citation linkages generate reliable indications of subject similarity and semantic relations among papers. Changes in co-citation patterns reveal the evolution of scientific disciplines and specialties over time, as well as their core literature (Small, 1973).

Co-citation patterns can change over time when earlier works receive additional citations. In contrast, the concept "bibliographic coupling" which depends on references listed in coupled documents is a fixed value without these changes (Small, 1973). The idea of bibliographic coupling was introduced by Fano (1956), but was made popular by Kessler (1963). Bibliographic coupling "happens when a reference is used by two papers as a unity of coupling between those two papers. The strength of bibliographic coupling depends on the number of references the two papers have in common" (Osareh, 1996, p. 155). According to Egghe and Rousseau (2002, p. 349) "two documents are said to be co-cited when they both appear in the reference list of a third document". The main difference between co-citation and bibliographic coupling is that the latter links source documents, whereas the former links cited documents (Small, 1973).

Co-citation has been widely used since its introduction (Boyack & Klavans, 2010) which indicates its importance in the bibliometric community. For example, Colavizza, Boyack, van Eck, and Waltman (2017) investigated the similarity of publication pairs at different co-citation levels: journal, article, section, paragraph, sentence, and bracket (including the citations). They found that the similarity of publication pairs increases monotonically as the co-citation level gets lower (from journal to bracket). "In other words, the lower the level at which two articles are co-cited, the more similar the articles will be on average" (Colavizza et al., 2017, p. 608).

### 6.1.2 Qualitative analysis based on the email in Appendix 8.6

Henry Small writes that the references in his paper really don't tell the full story of how the method of co-citation was developed. The idea of co-citation analysis comes from two fields (history of science and information science), but this was not apparent from the references which are mostly in information science. He emphasizes that the development of the idea is not reflected in the cited references, but in his work on the history of nuclear physics. The references are partly misleading because, for example, the two physics references in his paper were cited to explain the history of some topics.

Henry Small came from another field (history of science) and the idea emerged in his previous job at AIP's Niels Bohr Library in New York, when he was working on a project to document the history of nuclear physics. He was figuring out what nuclear physics was back in the 1920s and 1930s, who were the important people and what they discovered. He writes that he was doing bibliometrics without really knowing it. He did "pairings" of the different data elements: subject headings, authors, key words, references, etc. He brought his idea to bibliometrics when he started his new job in ISI in 1972. Working with bibliometric data, the idea was refined, and eventually found strong patterns of what he then dubbed co-citation.

## 6.2    h-index

### 6.2.1    Explanation of the concept

In 2005, Hirsch (2005), a physicist at the University of California, San Diego, proposed the h-index, an index for quantifying and evaluating individuals' scientific research output which had an enormous influence, not only on the bibliometric community (Waltman & Van Eck, 2012). "A scientist has index $h$ if $h$ of his or her $N_p$ papers have at least $h$ citations each and the other $(N_p - h)$ papers have $\leq h$ citations each" (Hirsch, 2005, p. 16569). For example, an h-index of 20 means that the scientist has published 20 papers that each received at least 40 citations. He indicates that a scholar's h-index increases approximately linearly over time, and cannot decrease with time. The "h-index was proposed as a better alternative to other bibliometric indicators (such as number of publications, average number of citations, and sum of all citations)" (Bornmann, Mutz, Hug, & Daniel, 2011, p. 346). Many empirical studies have used the h-index for ranking scientists in different fields. For example, Cronin and Meho (2006) used the h-index to rank influential information scientists. Besides scientists, the h-index has been used for ranking institutions (Mahmudi, Tahamtan, Sedghi, & Roudbari, 2015; Mitra, 2006) and journals (Braun, Glänzel, & Schubert, 2006).

The introduction of the h-index attracted enormous attention in the bibliometric literature, which resulted in numerous variants of the h-index (Bornmann et al., 2011). These variants were intended to eliminate some of the weaknesses of the h-index. The A-index (Jin, 2006) and AR-index (Jin, 2007) were among the first variants introduced. The meta-analysis of Bornmann et al. (2011) indicated however that a high correlation exists between the h-index and its variants (0.8 to 0.9). This means that there is redundancy between the original and its variants, and that the h-index variants hardly provide additional information to the original.

While average citation rates are likely to be inflated by a small number of highly cited papers, the h-index doesn't have this disadvantage (Hirsch, 2005). However, the h-index has many drawbacks. For example, "it is field-dependent, it may be influenced by self-citations, it does not consider multi-authorship, it is dependent on the scientific age of a scientist, it can never decrease, and it is only weakly sensitive to the highly cited papers" (Bornmann et al., 2011, p. 347). In addition, some previous studies have pointed to the arbitrariness of the definition of the h-index (see Ellison, 2013; Lehmann, Jackson, & Lautrup, 2008; Waltman & Van Eck, 2012). Lehmann et al. (2008) maintain that the problem with the h-index is that it assumes an equality between citation and publication counts; however, they are two quantities with different units. Waltman and Van Eck (2012) refer to the inconsistency of the h-index for measuring the overall impact of a scientist: the index "violates the following property: If two scientists achieve the same relative performance improvement, then their ranking relative to each other should remain unchanged" (p. 409).

### 6.2.2  Qualitative analysis based on the email in Appendix 8.7

Jorge Hirsch notes that he was not working in the field of bibliometrics and was unfamiliar with the literature. He states that the cited literature did not play an important role in developing the idea of the h-index. However, he refers in his text to his interest in citation impact as an objective way to evaluate the importance of research. Jorge Hirsch came to the idea of the h-index about two years before the publication of his idea in 2005. He came to the idea from a practical point of view (not from the literature), while evaluating applications for positions in his department. The index appeared as a simple way to summarize two different information items contained in citation records. He started using it and studying its properties but wasn't thinking about publishing a paper on it. The interest of and discussion with colleagues motivated him. The reading of the paper by Redner (2004) from the area of bibliometrics stimulated his interest in bibliometrics and the intention to publish the h-index idea.

## 6.1 Altmetrics

### 6.1.1 Explanation of the concept

The term alternative metrics (altmetrics) was first introduced by Priem, Taraborelli, Groth, and Neylon (2010). The term covers a group of metrics (e.g. Twitter and Facebook counts) which were proposed as an alternative to already existing metrics such as citation counts. Altmetrics became very popular right after their introduction. Altmetrics mostly refer to web-based metrics for measuring the scientific and public impact of publications (Bornmann, 2014). Altmetrics measurements are derived from the social web, including Twitter, Facebook, Mendeley, blogs, etc. (Thelwall, Haustein, Larivière, & Sugimoto, 2013). Impact is measured based on the number of times a paper has been downloaded (https://www.projectcounter.org), saved in CiteULik (http://www.citeulike.org) or Mendeley (https://www.mendeley.com), recommended (https://f1000.com/prime), discussed (https://twitter.com), etc. (Bornmann, 2014). The company Altmetric (www.altmetric.com) is one of the major providers collecting and analyzing altmetrics data (Tahamtan & Bornmann, 2018). The tool PlumX Metrics "provide[s] insights into the ways people interact with individual pieces of research output (articles, conference proceedings, book chapters, and many more) in the online environment"
([https://plumanalytics.com/learn/about-metrics/](https://plumanalytics.com/learn/about-metrics/)).

A recent review discusses empirical studies on altmetrics, its various interpretations, and its data collection and methodological limitations (Sugimoto, Work, Larivière, & Haustein, 2017). Advantages and disadvantages of altmetrics are reviewed in Bornmann (2014): advantages are possible measurements of broader research impact, the diversity in kinds of data and data sources (e.g. tweets, reads on Mendeley), and impact measurements shortly after publication. The most important disadvantage is that it is not clear what is being measured (valuable impact or background noise), while with citation counts it is obvious that we are measuring (valuable) impact on science. If altmetrics data is used for research evaluation, it should be normalized since newer papers and papers on certain topics (e.g. medicine) receive higher scores than older papers and papers on other topics (e.g. mathematics). While citations are normalized to allow cross-field comparisons, normalizing altmetrics data is still not common (Bornmann, 2014).

Previous studies indicate different correlations between altmetrics and citation counts (Costas, Zahedi, & Wouters, 2015; Thelwall et al., 2013; Zahedi, Costas, & Wouters, 2014). As the results of a meta-analysis show, "the correlation with traditional citations for micro-blogging counts is negligible (pooled r = 0.003), for blog counts it is small (pooled r = 0.12) and for bookmark counts from online reference managers, medium to large (CiteULike pooled r = 0.23; Mendeley pooled r = 0.51)" (Bornmann, 2015, p. 1123). This might indicate that some altmetrics are merely early indicators of subsequent citation rates (Thelwall et al., 2013), and that some altmetrics do not reflect the same kind of impact as citation counts (Costas et al., 2015). Mohammadi and Thelwall (2014) further show differences in the correlations between Mendeley readership counts and citations for different disciplines underlining the field dependency of these counts. Costas et al. (2015, p. 2003) find that "altmetric scores (particularly mentions in blogs) are able to identify highly-cited publications with higher levels of precision than journal citation scores (JCS), but they have a lower level of recall".

### 6.1.2   Qualitative analysis based on the email in Appendix 8.8

Before working in the field of information science, Jason Priem used to work as a middle school teacher in the early 2000s. At that time, he got interested in, and was fascinated by a growing "edu-blogosphere" of teachers with blogs, and always wanted to do something with studying online communities. He produced FeedVis, which provided a visualization of the edu-blogosphere that aimed to capture what the teachers were talking about at any given moment, and track how that conversation changed over time. Jason Priem learned about bibliometrics when he started his Ph.D. at the University of North Carolina (UNC). He wanted to do the same thing he had been trying to do with FeedVis, but instead of understanding the conversational topology of some blog network, he was trying to understand and map the frontiers of human knowledge. He was not that excited about doing more work on citations.

Thus, he went looking for other ways to map or analyze the scholarly conversation. At first, he spent a lot of time looking into argument mapping, work that was trying to construct networks of claims and evidence rather than networks of papers. He was interested in the idea of a network of claims and counter-claims that would look very much like a citation graph. He lost interest in this approach, however, based on his belief that manual annotation would not scale well and automatic annotation techniques were not yet mature. He writes that the PLOS ALM project, a new instrument with which to observe scholarly communication, was the main source which inspired him. For this, he gives enormous credit to PLOS for leading the way on altmetrics.

He started collecting some data using the PLOS ALM dataset, and started looking at the literature to see who else was talking about this. He read a lot of Blaise Cronin's work; his thought that there was a future in using the web to find scientific "street cred" (Cronin, 2001) gave him a lot of encouragement. In addition, he was inspired by Mike Thelwall's webometrics work. Jason Priem writes that both Blaise Cronin and Mike Thelwall were way ahead of him.

He believes that Blaise Cronin and Mike Thelwall had the ideas behind altmetrics before him, when the online ecosystem hadn't matured enough to fully support what they were trying to do. His discussion with Cassidy Sugimoto (who at that time was a doctoral student) encouraged him too.

He came up with the name "altmetrics" while in the shower and tweeted it straight after to see if that name would get any traction. He used to think that there was no way the bibliometrics community would ever accept the name he had chosen for his indicator – altmetrics. So he came up with another name, and published an article in First Monday about "scientometrics 2.0" (Priem & Hemminger, 2010). He felt the idea needed a real strong publicity push so that other people would get excited about it and start doing the research and building the tools. Therefore, he wrote and published a "manifesto" about the idea. He got the idea of writing a manifesto from his background in history and the 20th century art movements when many of these manifestoes were published. He thought that no one would listen to a PhD student. Therefore, he realized that having some coauthors that people respected would influence the publicity and acceptance of the idea. He wrote the manifesto when attending an ASIST conference, in between sessions, running back and forth up to his hotel room. He invited three coauthors, none of whom were information scientists, but all of whom had written altmetrics papers of some kind before. They gave him some great thoughts on the manuscript and how best to present it. He believed that SILS had a culture that expected graduate students to be contributing to the field, which also motivated him in building upon his idea.

## 7 Discussion

In this case study, we examined whether creativity of papers can be measured on the basis of cited references (see Uzzi et al., 2013; e.g. Wang et al., 2017). Although this question has already been studied in patents (Callaert et al., 2014), we found no study exploring the roots of creativity in papers. We interviewed the corresponding authors of eight breakthrough papers in the field of scientometrics. They were asked to indicate the roots of their creative ideas, and whether these creative ideas were rooted in the cited references in their breakthrough papers. In the previous sections, we presented the detailed interpretation of the emails which were sent us explaining the roots (the emails themselves are in the appendix). The emails are not only interesting because they explain the link between cited references and ideas, but also because they personally explain the origin of ideas for important concepts in scientometrics.

In rare cases, authors of the landmark papers referred to some cited references which had inspired the development of their creative ideas, where some references are more important than others. Even in patents, most scientific references are not important for the development of the invention, perhaps being used as background information (Callaert et al., 2014). Although new ideas in science can be inspired by previous studies (Zeng et al., 2017), this seems not to be the case in all breakthrough papers in scientometrics. Michel Zitt emphasized more than others the importance of the literature in the development of citing-side normalization. He writes that the references of their 2008 article are critical to the emergence of citing-side normalization or fractionated citations. However, Zitt and Small (2008) have mainly used traditional scientometrics literature. Michel Zitt doesn't refer to any study outside scientometrics as the basis of citing-side normalization (which would lead to uncommon combinations of cited references). Another concept which was mostly on the basis of previous scientometrics research was Henk Moed's crown indicator (see section 6.4). Unlike the study of Uzzi et al. (2013), conventional combinations of prior work and incorporating atypical combinations do not seem to be related to creative ideas in scientometrics. In addition, our findings revealed that novelty in scientometrics (breakthrough papers) does not reflect "the recombination of pre-existing knowledge components in an unprecedented fashion" (Wang et al., 2017, p. 1417). In our later contacts with Henry Small, he said that there are some discoveries that are the direct outcome of combining ideas in cited references, but not many.

Publications might be cited in the landmark papers without being critical for the development of the idea. For example, although the reading of a certain paper from the area of bibliometrics stimulated the interest of Jorge Hirsch (who introduced the h-index) in bibliometrics and the idea to publish it, the publications cited did not play a major role in the development of the h-index. Wolfgang Glänzel notes that because of the relationship with the characterization theorem published in Glänzel, Telcs, and Schubert (1984), he called the newly developed method "Characteristic Scores and Scales". However, the 1984 paper is not cited in the breakthrough paper. Jason Priem refers to the importance of the works of Blaise Cronin and Mike Thelwall, but the works of these scholars are not cited in Jason Priem's manifesto. The interview of Callaert et al. (2014) with 33 Belgian inventors also indicated that non-patent references (scientific references) are the source of inspiration for only about 50% of the inventions. However, 30% of the patents that were inspired by scientific references did not contain any scientific references. Nagaoka and Yamauchi (2015) also indicated that only half of the patents relied on scientific sources in their Research & Development processes, among which only 17% revealed such important literature in their patent documents. It seems that publications might be read but not cited for certain reasons (see Tahamtan & Bornmann, 2018).

Our study revealed that co-citation analysis and the h-index are not rooted in the scientometric literature. This finding challenges the claim that novel papers use a combination of traditional and atypical references (see Uzzi et al., 2013). According to Uzzi et al. (2013, p. 471) "creativity in science appears to be a nearly universal phenomenon of two extremes. At one extreme is conventionality and at the other is novelty". A similar approach was used by Veugelers and Wang (2016); Wang et al. (2017) found that highly novel papers are those that make more distinct new journal combinations in their reference lists.

The failure to cite publications which played an important role in the development of certain ideas might have other than intellectual or cognitive causes. Some journals determine "the number of references that should be used for different types of papers in their guidelines or instructions for authors. For instance, based on the 'manuscript formatting guide' of *Nature*, articles (original reports) should have no more than 50 references" (Tahamtan & Bornmann, 2018, p. 211). In this regard, Michel Zitt notes that he and his coauthors were limited by the rules on bibliography length in the *Journal of the Association for Information Science and Technology* (JASIST) and they had to sacrifice some references. András Schubert notes that because of the strict space limitations given by the 1983 Varna conference, no references were allowed in the proceedings material.

Thus, the results of this study indicate that creative ideas might rest partly on the shoulders of past research (which is cited in the landmark papers). Even in inventions (patents) scientific literature might not play a significant role in the development of ideas (Nagaoka & Yamauchi, 2015). Our results further show that other sources of creativity exist, one of which is the need to find a solution to a practical problem. Some ideas are initially used in practice before they have been published. For example, Jorge Hirsch came to the idea of the h-index while evaluating applications for positions in his department as a simple way to summarize the information contained in citation records. He did not even think about publishing this idea. The co-citation idea arose when Henry Small was working on a project to document the history of nuclear physics in his previous job at AIP's Niels Bohr Library in New York. The RCR indicator was developed in a similar situation.

We note that 'biosociation' (Koestler, 1964) is the principle on which the technique of measuring creativity by using pairs of cited references is based. However, this technique only works if the idea is ready-made (see above). In the case of custom-made ideas this does not work (see Salganik, 2017). Ready-mades are breakthroughs which are rooted in existing literature. Custom-mades are those which are not based on previous research. In our study, the crown indicator could be considered as ready-made because it is somewhat based on previously developed indicators. Co-citation analysis is custom-made, because it has been developed out of the blue.

Being aware of the prior literature is a key factor in citing relevant papers, said Henry Small. He referred to the difficulty of searching the relevant literature in the past. In our further email exchanges with Henry Small, he mentioned that he had trouble coming up with a reference list for his co-citation paper, wandering around the Drexel University library trying to find something to cite. He noted that some years later he did find a highly relevant prior study by an information scientist/linguist named Bar-Hillel in which a similar idea was called 'co-quotation' (Bar-Hillel, 1957). He said that if he had known how to search the literature better, perhaps he would have uncovered and cited this relevant paper. However, when he wrote his first paper on co-citation there was not a lot of literature classified as bibliometrics. Therefore, some scientists may not have had much awareness of the prior literature when they came up with their creative ideas. Since this has been significantly changed and the relevant literature can be found in the literature databases today, a time effect should be visible. Further studies should investigate whether the ideas for more recent landmark papers than the old ones are better rooted in the literature (and cite prior ideas which are similar).

One might expect that breakthrough ideas in a field are developed by specialists with long experience. Our results show, however, that several ideas have been developed by outsiders to the field of scientometrics. Some landmark papers were published when their authors were not very prominent in the field. For example, Henry Small published his idea when he was not familiar with the literature and history of information science. The main basis for altmetrics has been Jason Priem's background and experience in following the online community of teachers, which made him think about measuring science with online data rather than citations. One reason could be that people with backgrounds other than the area in which they contribute to (e.g. scientometrics) think about what they do outside the box. However, other cases are in agreement with the expectations, i.e. experts in the field proposed the creative ideas. For example, Wolfgang Glänzel developed the CSS method as a professional bibliometrician. In some cases, ideas emerged when the interviewed author was not working on a particular scientometric problem; however, working with scientometric data helped them in refining and improving those ideas (e.g. the concept of co-citations).

Among the literature, Ph.D. dissertations seem to be important for the development of the CSS method and top decile performance indicator. For example, Wolfgang Glänzel refers to his doctoral dissertation (1984) in mathematics, which has been related to the modeling of stochastic processes in bibliometrics.

The interest of and discussion with colleagues seem to be an important contextual factor for the development of creative ideas in scientometrics. Interest among colleagues who take up the ideas and good networking among colleagues are prerequisites for scientists to make crucial contributions to scientific revolutions (Bornmann & Marx, 2012). Interacting with peers is a powerful tool in organizations for the emergence of creativity (e.g. Erren et al., 2017; Neumann, 2007). Creative scientists more than other scientists discuss with their colleagues, as valuable sources of information (Kasperson, 1978). Correspondingly, colleagues have had an important role in the development of breakthrough papers in scientometrics. For example, Henry Small's talk to and collaboration with information science people such as Belver Griffith at Drexel University, Jorge Hirsch's and András Schubert's talk with their colleagues played critical roles on how to proceed with their ideas. Although interacting with peers matters, what seems to be common among the authors of breakthrough papers is their internal creativity. Individuals' cognitive and affective features might influence creativity, for example, by facilitating intrinsic motivation, flexible thinking and problem solving (Hennessey & Amabile, 2010).

Another important factor in the development of creative ideas is personal interest in certain areas of research, and this was clear in all participants' responses to our survey. Wolfgang Glänzel writes that one basis for his landmark paper was his personal interest in probability distributions; Jason Priem's interest in online communities inspired him to pursue his studies in a field that could help him investigate data on social networks. It seems that researchers have to be fascinated by a specific topic or a certain area to develop important core concepts in a field. The fascination might refer to an area outside the field of scientometrics as in the case of Jason Priem or inside the field as in the case of Wolfgang Glänzel. In addition, thinking about doing something different could also lead to innovations, alongside many other factors. For example, Jason Priem thought about measuring science using something other than citations, or Henry Small was thinking of doing something that no one had done before. That's also an important motivator, the drive to be different.

Technology and instruments are also important factors in developing some ideas. Yaqub (2018) writes that observations are usually mediated by the development and use of instruments. For example, technology developments in the field of scientometrics lead to a professional and wider application of the RCR method. This has become clear from András Schubert's response to our survey. He writes that a big breakthrough in the early 1980s was the possibility of using the magnetic tapes of the SCI and processing them by computer. Henry Small points to the importance of the SCI. He refers to Irina Marshakova who had published her article (in Russian) on the same idea at about the same time (Marshakova, 1973), yet had no way to test her idea, while Henry Small who worked at ISI could automate the process using SCI. Wolfgang Glänzel notes that the robustness and the real advantages of CSS became apparently twenty years later, when hardware and software development and database access allowed broadening of the scope beyond journal analysis and application of the method to larger citation windows.

Our study indicates that some creative ideas were published a couple of years after they were developed. Some ideas might also get rejected probably because they are ahead of their time, or they might get rejected in one field but be welcomed in other fields. For example, Henry Small noted that one of the reasons why his co-citation paper was rejected by a history of science journal was that it was too statistical, and maybe historians didn't want that sort of thing. However, delay in ideas gaining recognition is not something new or unusual. For example, Becattini et al. (2014, p. 1) note that "the 2013 Nobel Prize in Physics was awarded to Higgs and Englert for their prediction of the existence of the Higgs boson. Though the Higgs particle was experimentally discovered at CERN in 2012, the original theoretical works date back to the 1960s. Thus, it took about half a century of intense work to confirm their prediction". Sometimes new ideas are at first not followed up by the scientific community, for example, because they are difficult for colleagues in the field to interpret or are not in line with an already existing discussion (e.g., the core questions in a field) (Bornmann & Marx, 2012). CSS was among the concepts in scientometrics that passed into oblivion for almost twenty years, while altmetrics was taken up right after its publication.

Another finding of our study is that ideas are sometimes named after being published, or the names of indicators have changed after concepts have been published. For example, the idea of András Schubert was first named 'relative impact' in Schubert, Zsindely, Glänzel, and Braun (1982) and later 'relative citation rate' in Schubert et al. (1983), or the first label for co-citation was "pairing". Today, RCR is mainly known as the mean normalized citation rate (CWTS) or field weighted cited rate (Scopus); the top decile performance indicator is named today PP(top 10%) or excellence indicator. The abbreviation RCR is used curiously enough for a recently introduced field-normalized indicator by the NIH – the Relative Citation Ratio (Hutchins, Yuan, Anderson, & Santangelo, 2016). Giving attractive names to new paradigms in a field makes it easier for them to be understood and established (Bornmann & Marx, 2012), such as altmetrics in scientometrics. Here, an interesting question is why the name altmetrics has been established for this group of new metrics. Other names have been proposed, such as article-level metrics (Neylon & Wu, 2009), influmetrics (Cronin & Weaver, 1995), etc. What are the driving factors for one direction rather than a possible other direction? Another interesting point is that some ideas published in landmark papers have been published earlier or at the same time by other people, and maybe different names have been used for these ideas. Which factors lead to the status of a landmark paper in one case, but not in the other case? Is the Matthew effect active here? Merton (1963) noted that many similar scientific discoveries and inventions in science have been made by scientists independently of one another and more or less simultaneously. He also mentions that "sometimes a scientist will make a new discovery which, unknown to him, somebody else had made years before" (Merton, 1963, p. 237). However, only some of these discoveries and breakthrough papers gain recognition by the scientific community.

Some ideas in scientometrics have developed over time. For instance, Michel Zitt mentions that his 2008 paper (Zitt & Small, 2008) did not include a formal demonstration of the new indicator's properties, which were studied in subsequent articles, such as Zitt (2011) and Waltman and van Eck (2010). Another example is RCR. In András Schubert's opinion, RCR was already presented in several publications, such as Schubert, Zsindely, and Braun (1985) and gained its final form and place in Schubert and Braun (1986). In addition, a short introduction of CSS was published in *Scientometrics* (Schubert, Glänzel, & Braun, 1987b), and the theoretical framework explaining the methodology and providing more detailed application was published in the *Journal of Information Science* (Glänzel & Schubert, 1988).

It is difficult to arrive at a general conclusion based on our results, because we studied the papers from only one field and also a limited number of breakthrough papers. Therefore, the results are not generalizable (to other disciplines). However, our results reveal that further (comprehensive) studies are required to investigate the question.

In conclusion, our survey shows that past published research is not the main root of creative ideas. Some published research might be important for the development of the creative idea but might not necessarily be cited in the breakthrough paper. Also, the cited references might be related to the breakthrough paper; however, the creative ideas haven't necessarily been inspired by them. The references have been included in the breakthrough paper for embedding the paper in the corresponding scientometric context. We didn't find any source of serendipity in the emails from the authors of the landmark papers. Most creative ideas in scientometrics have been the result of finding solutions for a given (practical) problem. Networking and discussion with colleagues is another main source for the inspiration and development of creative ideas.

It is possible that atypical combinations of factors, such as solutions for practical problems, contacts, informal communications, are important sources of creativity and novelty rather than atypical combination of cited references. Whether this is really the case should be examined in future studies.

# Acknowledgements

We thank all authors of the landmark papers in scientometrics for their time and efforts in supporting our study.

# 8    Appendix

We had some other email exchanges with the authors of the breakthrough papers which are not mentioned in the appendix, yet, they are rather included in the discussion section.

## 8.1    Appendix for András Schubert's email regarding Relative Citation Rate (RCR)

Dear Lutz,

As the saying goes, I haven't got time enough to tell the story in short, so please forgive me, if I'll be a bit longer than expected. The idea of RCR came to my mind much earlier than I became a "professional" in scientometrics. I was introduced to scientometrics by the late Imre Ruff, a fellow professor of chemistry to Tibor Braun at the Department of Inorganic and Analytical Chemistry of the Roland Eötvös University of Budapest. Imre was one of the reviewers of my book on reaction kinetics in the mid-70s, and this was "the beginning of a beautiful friendship", which was ended only by his untimely death.

The focus of our discussions has gradually been shifted from physical chemistry to scientometrics, particularly, when Imre decided not to join the newly organized information science and scientometrics unit at the Library of the Hungarian Academy of Sciences lead by Tibor Braun because of a very challenging new topic he found in his chemical research, and he offered me the possibility to apply for the position to be the leader of the scientometrics group. Later, I accepted this offer (and, using the platitudinous cliché, the rest is history...), but at the time of the story I was just tasting the topic.

One of the evergreen questions was the possibility of cross-field comparisons by citation indicators. I remember quite sharply that I waited for Imre in the library room of the Department (actually, a small and dark vestibule of the laboratory), when I devised the idea of measuring the relative contribution of certain "actors" (countries, institutions, departments) to the impact factor of some key journals.

This might have resulted in a field-independent indicator of citation eminence. Since at that time we only had the printed volumes of Science Citation Index, to put the idea into practice seemed to be quite a laborious task. Imre was rather enthusiastic about the idea, and gave some advices how to proceed. We named the new indicator "relative impact". I spent many hours with tallying citations until we had the first results, and they were quite promising. We did not think about publishing our findings (the journal Scientometrics has not even been launched then), but we kept it in our toolbox for internal use.

When I began my work as the leader of the scientometrics group, I had to produce a plenty of practical evaluation reports and comparative studies. The Hungarian Academy of Sciences (HAS) had a network of almost 200 research units (standalone institutes and supported groups located in universities), and the leaders of the Academy were eager to know their relative standing. Another key question was the comparison of the performance of the Academy with that of the universities. (How this information was used and misused, that's another story.) The relative impact had a key role in these analyses. These studies were not public, they were confidential reports to the leaders of the academy.

To my best knowledge, the first publication where the "relative impact" was used was the volume "A tudományos publikációs tevékenység mutatószámai az MTA természettudományi, műszaki, orvostudományi, és agrártudományi kutatóhelyein, 1976-1980" (Indicators of scientific publication activity in the natural science, engineering, medical science and agricultural research units of the HAS, 1976-1980) published in 1982 by the Library of the HAS as Volume 2 of the series "Informatics and Scientometrics". No author was indicated on the cover, the inside title page listed the "compilers" of the volume: András Schubert, Sándor Zsindely, Wolfgang Glänzel, Tibor Braun. Because of the technical difficulties in manually tallying the citations, only the three most significant journals were taken into account for each institution.

A big breakthrough in the early 80s was the possibility of using the magnetic tapes of the SCI and processing them by computer. We could analyze volumes of data unimaginable earlier. The results of the first large-scale international comparison of scientometric indicators were published first in Hungarian as Volume 3 of the above mentioned series of our Library: Schubert András, Glänzel Wolfgang, Braun Tibor: Tudománymetriai mutatószámok 32 ország természettudományos alapkutatásának összehasonlító elemzéséhez, MTAK, 1983. Later it was published in English as T. Braun, W. Glänzel, A. Schubert (1985) A 32-Country Comparative Evaluation of Publishing Performance and Citation Impact. Singapore; Philadelphia: World Scientific, 1985. vii, 424 p. https://doi.org/10.1142/9789814415149_0005. Wolfgang's main contribution was the addition of a statistical reliability estimation to the bare indicator and, of course, as programmers in emergency, we did all the programming work (in FORTRAN – the only programming language we knew) together.

We presented the indicator as a new methodological tool with Wolfgang in the 1983 Varna conference. That was the first time we used the term "Relative Citation Rate" for the indicator. Because of the strict space limitations, no references were allowed in the proceedings material.

The RCR were used in 1985 in the paper A. Schubert, S. Zsindely, T. Braun: Scientometric indicators for evaluating medical research output of mid-size countries. Scientometrics, Vol. 7, Nos 3-6 (1985) 155-163.

In my opinion, it gained its final form and place in: A. Schubert, T. Braun: Relative indicators and relational charts for comparative assessment of publication output and citation impact. Scientometrics, Vol. 9, Nos 5-6 (1986) 281-291.

Best regards, András

## 8.2 Appendix for Wolfgang Glänzel's email regarding Characteristic Scores and Scales (CSS) method

Dear Lutz,

As promised in my previous mail, I send you further explanation regarding the origin of the CSS method. Although this was 30 years ago when this idea emerged and I do not recall all details and steps in the development of this methods, I still remember the main "sources" and ideas that finally resulted in detecting the mathematical background of the method.

One basis was the analysis of citation distributions of journals. In the early 1980s we compiled the citation distributions of all journals covered in the SCI database. We observed that, while the mean citation rate of journals (e.g., their Impact Factors) could be similar, their shares of cited papers (or equivalently, their uncited papers) was sometimes quite different. We have therefore published the shares of uncited papers along with the journal impact measure (cf. Schubert & Glänzel, Scientometrics, Vol. 5, No. 1 (1983) 59-74). At that time, we also experimented with using the citation impact of cited papers and other conditional citation means as a supplementary journal citation measures. This was also closely connected with my personal research interest in probability theory that was focused on the analysis of the probability distribution. At that time I was working on my doctoral dissertation in mathematics related to the modeling of stochastic processes in bibliometrics (1984). The analysis of citation distributions revealed a remarkable property, namely, that the mean values of truncated distributions approximately form a linear function of the points, at which they are truncated. This is a characteristic property of power-law distributions. A theoretical generalization of these observations resulted in one of the main findings of my dissertation and the publication of the article by Glänzel et al., Z. Wahrscheinlichkeitstheorie verw. Gebiete, Vol. 66, No. 2 (1984) 173-183. In my opinion, this article can be considered the main theoretical basis without which the CSS method had probably never been developed, systematically studied and applied in its present form.

Unfortunately, I do not remember how the second idea emerged: the idea of truncating the distribution at their mean value and of repeating this procedure on the iteratively truncated samples. This procedure can be conceived as creating a series of specific conditional citation means. Because of the relationship with the characterisation theorem published in Glänzel et al. (1984), we called this method "Characteristic Scores and Scales". We first used the method for journal analysis and its very first version still used the auxiliary class of uncited papers, which is not truly related with the method. Later we have dropped this class.

The proposed method has been published in two articles, a short introduction with a sample illustration as a "World Flash on Basic Research" in Scientometrics (Schubert et al., 1987) and the theoretical framework explaining the methodology and providing more detailed application in Journal of Information Science (Glänzel and Schubert, 1988). In this article, we already studied the basic mathematical properties of the method. Obviously, the above mentioned linear function associated with power-law distributions would yield a geometric serious for the CSS model only depending on the tail parameter of the distribution (and its expected value, of course).

Despite its interesting properties, the method passed somewhat into oblivion. This was probably due the enormous computational effort and CPU-time that was required to compile the largescale scores and classes using the main-frame computer infrastructure of the 1980s. The robustness and the real advantages of the method grew apparent only twenty years later, when hard- and software development and database access allowed to broaden the scope beyond journal analysis and to apply the method also to larger citation windows (even 20 years or more), to multidisciplinary environments and to various level of aggregations such as to countries, institutions and individual scientists.

Best regards,

Wolfgang

*The second CSS source paper*

Glänzel, W., Schubert, A., Characteristic Scores and Scales in assessing citation impact. Journal of Information Science, 14 (2), 1988, 123-127.

## 8.3 Appendix for Francis Narin's email regarding top-decile performance

Dear Lutz,

Sorry to be so slow in responding to your letter, but I have been busy with computer problems, and trying to find a copy of reference 2, in which I know we used top decile citation data. Unfortunately I cannot find a copy, and don't know anybody currently in the evaluation office at NIH [National Institutes of Health], and I doubt if they would be able to find a copy of a report from 1983 anyway.

That report however was based almost entirely on my doctoral dissertation, which was performed with the cooperation of NIH and with an NIH based data set, so I am enclosing the section of my doctoral dissertation (reference 1), which contains a discussion of the data and its significance in the NIH context.

The problem we faced in looking at the citation impact of the papers, both intramural and externally supported, of the various institutes at NIH was that their subject areas within biomedicine were radically different, and contained papers concentrated in medical areas for which the citation densities were radically different. For example, NIGMS [National Institute of General Medical Sciences] is primarily a basic medical research Institute, publishing largely in the biochemical areas of very high citation. In contrast NIDR [National Institute of Dental Research] primarily supports and performs dental research, a clinical area in which the citation density is far smaller than biochemistry.

As a result we had to devise an indicator of research excellence that was independent of the citation frequency differential between different areas of biomedicine, and the way we thought of doing this was by looking at the top 10% of most highly cited papers, the top decile, and comparing the different institutes in terms of the percent of their papers in the top decile, their top decile performance in these specific areas, essentially normalizing the Institute data for research area.

You will note in the enclosed section of the dissertation that all institutes, and all the paper supported by the institutes, have remarkably good citation data in the areas in which they concentrated. We did note that NIDR supported papers published outside of their core area of dental research were not especially highly cited, which can be explained away if we assume that more basic research that is important to dental research may not be of universal biomedical research importance.

Overall however the top decile performance of all of the individual institutes at the NIH, typically 20% or more, was quite remarkably good in the 1980s.

One technical clarification: the B/I/D database used in all of our work for NIH was a collection of all the papers in approximately 275 core biomedical journals which contained 75 to 80% of all the papers in which NIH intramural scientists and NIH grantees publish. As part of CHI Research's contracts with NIH we manually looked up, in the various university libraries in the Philadelphia area, all the papers in these 275 journals in some 10 years or so, and recorded the research support acknowledgments in the papers, producing a cited set of papers for which we knew both the institutional affiliations of the authors, and their sources of research support.

I hope the above explanation, and the attached extract from my dissertation, are enough to pinpoint the origin of our top decile indicator.

If you need any more information, please feel free to contact me.

Francis

2. "Subjective vs. Bibliometric Assessment of Biomedical Research Publications", U.S. Department of Health and Human Services, Public Health Service, National Institutes of Health, April 1983.

1. "Concordance between Subjective and Bibliometric Indicators of the Nature and Quality of Performed Biomedical Research", Doctoral dissertation, Walden University, April 1981.

## 8.4 Appendix for Henk Moed's email regarding crown indicator (CPP/FCSm)

Dear Lutz,

This is an interesting project you are working on. Here are some notes on my perception of the degree of originality of our indicator, and on why we cited particular papers in our paper.

The idea to compare the citation impact of a unit of assessment (UoA, e.g., author, group) with that of a particular benchmark set, and calculate the ratio between the two citation rates, was not new. The ISSRU [Information Science and Scientometrics Research Unit] group (Wolfgang, Braun) did that already in the early 1980s; they named their indicator relative citation rate (RCR) if I remember well. And also in the Leiden report "On the Measurement…." it was done.

There were various ways to define the benchmark set. ISSRU compared the average citation rate of a country's articles with that of all papers published in the journals in which that country had published- in the early CWTS terminology, a CPP/FCSm-type of indicator that was highly criticized by many researchers interviewed in validation rounds conducted at Leiden University in the early 1980s. Vinkler (1986) defined a relative citation rate using journal categories as benchmark sets.

In our 1995 paper we tried to generalize the various versions of the relative citation rate, introducing a citation balance model described in the paper. So it was logical to cite the work on which we built our version of the relative citation rate (Glanzel, Narin, Vinkler). Our symbol was CPP/FCSm.

The citation balance model is to the best of my understanding original. What is probably also original is that our relative citation rate did not only take into account (correct for differences in) the subject field (journal category), but also the year of publication, and the type of document. This model could be applied to any set of target articles under evaluation.

It was applied also to journals. A relative journal impact ratio was calculated, taking into account the age distribution of papers, type of documents published, and the subject fields covered, JCSm/FCSm.

But the idea to compare a journal's citation rate with a wider field average was not new either. Some authors may have done this already before us, I do not remember this. A Russian female colleague, Irina Marshakova, proposed it in a 1996 paper, but it was said that she had already done this many years earlier in a Russian paper.

Our method was developed I think already in 1990 or so, and was used in several CWTS studies published in 1991 and later. Some of these works are cited as well in our 1995 paper. If I remember well, there were also a few "setting the stage" cited references. I could write a lot more about the considerations that we had in the development of this indicator, (including the issue as to why we decided to use in the citation balance model the "ratio of averages(or sums)" rather than the "average of ratios" in the model), but I am not sure whether it is relevant, and, in any case, I do not have the time now to do this. I hope this is enough for now.

Good luck with your interesting study!

Henk

## 8.5 Appendix for Michel Zitt's email regarding citing-side normalization (audience factor)

Dear Lutz,

The productivity of weak ties, distal transfers or multidisciplinary linkages has been quite a challenging topic for a long time, and I wish your team every success in this study. This is an attempt to sketch out the history of the particular paper you mentioned.

**1. The history of *'Modifying the Journal Impact Factor by Fractional Citation Weighting: The Audience Factor'*, according to our best knowledge.**

I'll try to distinguish the various tracks leading to Zitt-Small (2008). The references cited in its bibliography are in bold face. This article did not introduce the idea of citing-side normalization (fractionation), which already existed in three distinct research lines in the 1970s. Its aim was to characterize the features of the method as a general tool of citation normalization with a focus on journal impact.

**i. Forerunners.**

The influence of bibliography length was recognized quite early by ISI-Philadelphia team. Henry Small and Ed Sweeney, who applied fractional counting in cocitation indices (Small & Sweeney, 1985) refer to two independent forerunners: Tyler Thompson of Rutgers University and Martha Dean of ISI, who proposed in 1976, fractionation of citations to overcome the problem of bibliography length bias.

**ii. Influence weight**.

In the same pivotal year, CHI Research (F. Narin) published two citation classics: *Evaluative Bibliometrics: The Use of Publication and Citation Analysis in the Evaluation of Scientific Aciivity*, which stressed the problem of referencing, and **Pinski & Narin (1976)** *Citation Influence for Journal Aggregates of Scientific Publications: Theory with Application to the Literature of Physics*". This paper introduced *influence weights* with a very scarce reference list. We don't know if the team at CHI had some prior knowledge of Thompson or Dean, or not. It is also unclear to me if they knew of Yule and Herbert Simon's work which inspired Price's pathbreaking model. In the same year, 1976, Price proposed the dynamic view of citation networks generated by a "preferential attachment" (this later appellation coined by Barabasi & Albert), in other words a modelling of the Matthew effect.

The status of Pinski and Narin's work is of course distinct from Thompson-Dean's ideas. Firstly, it included a full development of the radical idea on iteration of influence. The process did include a treatment of bibliography length, but intricately linked it with the process of iteration.

Secondly, Pinski and Narin paper had a tremendous influence; in the next two decades of publication few replications were made due to difficulties of implementation at a large scale in those days, and also for reasons of tuning and interpretation. Later, it was put under the spotlight by Brin & Page (1998) as an inspiration, among other sources, of the Google algorithm. **Palacios-Huerta & Volij (2004)** and **Kodrzycky & Yu (2005)** reactivated Pinski and Narin's methods in journal evaluation, using new variants. PageRank as such was taken back to journal bibliometrics by **Bergstrom 2007** and **West et al. (2008)** (Eigenfactor) and **Moya-Anegon (2007)** (SJR).

**iii. Work on diversity and criticisms against conventional normalization methods.**

We might have started from an instantiation of Pinski and Narin on the grounds that it was not necessary to reinforce the Matthew effect implied by iteration of sources prestige, and just retained from Pinski and Narin, the built-in referencing correction and not the recursion. That was not the way; my personal experience was a bit different, and focused in practice towards efficient normalization methods, with a background in the issue of field diversity.

The recognition of diversity of citing behavior was the motivation of the forerunners quoted above. Narin's team and **Murugesan & Moravcsik (1978)**. Also **Garfield (e.g. 1979)** repeatedly stressed the referencing bias, rather than the size of fields, for example, as a determining factor of the impact level. We cited **Biglu (2008)** as testimony to the continuous flow of work in this line.

As my team became initiated to scientometrics on the mapping side, through French-style coword analysis (Callon, Courtial, Turner group at CSI-EM Paris, early 80s) along with ISI-style cocitation. We use a kind of fractional measure in our cocitation studies (**Zitt & Bassecoulard 1994,**1996) in the wake of Small and Sweeney. In *Scientometrics* (2003), we published an empirical bivariate characterization of both citing and cited data in a closed field, a kind of ideal-type of cocitation process modeling the maximum retrieval in function of thresholds both on the citing and the cited side. Later, I had some occasions to comment on bibliometric treatment of diversity (Measurement, 2005, ESEP 2008).

But the main driving force of my team's work - as I was involved in production of scientometric indicators at OST Paris - was the interpretation and limits of classical indicators of citation. The background of impact factor (or other citation indicators), and especially the question of field normalization is evoked in the citations to **Garfield (1972)**, **Sen (1992)**, **Marshakova (1996)**, **Schubert & Braun (1996)**, **Garfield (1998)**, **Bollen et al. (2006)**, **Fassoulaki et al. (2002)**, **Vinkler (2002)**, **Glanzel & Moed (2002), Pudovkin & Garfield (2004), Sombatsompop (2004), Rousseau (2005)**. Clearly, due to the expansive literature on the subject, this selection of references is adequate but not exhaustive.

Several works of ours addressed these questions, especially the classical cited-side normalization methods, which at this time, prevailed in evaluative scientometrics and still does perhaps. One of our works addressed the instability issue, when the zoom on the embedment of (sub)fields, with different citation behavior, varies (relativity of citation measures, **Zitt, Ramanana, Bassecoulard (2005)**: a kind of effect of the fractal structure of science on the stability of indicators. Remember that in these days, van Raan, Katz and Leydesdorff paid attention to the fractality of science networks. In the 2005 paper, we entitled a paragraph "cited-side or citing-side normalization", crediting **Small & Sweeney** along with a mention of **Thompson and Dean**. The same idea is more developed in the above-quoted Measurement (2005) (a comment of a general article on indicators by van Raan). This highlighted the dependence of relative impact measures on the definition of the basis of normalization, delineation/scale of fields. Depending on the options, changing the point of view towards citing-side, eliminates or alleviates this dependence towards field delineation (for example. the audience factor, using the journal-level granularity, is field-independent).

**iv. Audience factor, SNIP and other work on citing-side normalization.**

With these ideas in mind (and another project prolonging the question of instability on cluster embedment) and the necessity to implement it on convenient data, I convened with Henry Small at two conferences, Leiden STI 2006 and Madrid ISSI 2007. We cannot remember at which conference we discussed citing-side/fractionation matters. Anyway, our work in late 2006 and/or 2007 resulted in the publication of the 2008 article. In the meantime, Henry Small had attracted my attention to Bergstrom's work. This urged us to comment on the revival of Narin & Pinski filiation. Hence the recent references along this track, quoted above.

The 2008 article claims that there is room for this kind of partial normalization, which, as shown in its Table 3, does not ignore the position of actors, journals, etc. in the asymmetrical chains of knowledge (cell biology rated higher than medical applications), while correcting for referencing habits, without artificial reinforcement of hierarchies. Note that Zitt & Small (2008) did not include a formal demonstration of the properties, which were studied in further articles, especially: Zitt (2011) on the decomposition of the usual impact factor; Waltman & van Eck (2010) on the relations with influence/PageRank algorithms, detailing how the citing-side normalization is a limiting case of influence measure by removing recursion; Zitt & Cointet's (publication date) on macro-level properties of this outlinks-based normalization in a general directed-network weighting perspective (see also Zitt, JOI (2010). Although not completely finalized, the latter works (from 2013 on) depict how in the science system decomposed at some scale into entities (ex. fields), the variability of these entities' normalized impact over the system reflects its degree of multidisciplinarity (citation exchanges in the system), its structural change (differential growth of fields) and their interaction.

The 2008 article received some attention from the community, but neither Henry nor I developed it on an industrial scale. This was made along variants, mainly by CWTS and Elsevier-Scopus (Moed's SNIP and later variants by Waltman & van Eck). Glänzel, Leydesdorff and I also propose some variants.

## 2. Self-appraisal of Zitt and Small (2008) references.

I think that the references of the article are by and large faithful to the background of citing-side normalization/fractionated citations, even though in the text we might have been more specific about the construction of Pinski and Narin influence weights (Frank Havemann pointed out to me the interest of Geller's early contribution (1978) in terms of Markov chains). I must say that both Henry and I were reluctant towards Pinski and Narin iterations in a citation phenomenon already characterized by the Matthew effect. Among the technical advantages of the iterative approach is the automatic elimination of low-end literature, which can locally trap an uncorrected citing-side approach (low-end articles or journals with scarce bibliographies). I think Henry Small's background and mine were not that different, since I was familiar with pioneering ISI works on related topics, adding perhaps my own interests on the limitations of usual indicators towards the diversity of science.

To summarize, the text and the structure of the bibliography expressed that the citing-side/fractionation approach, encouraged by growing evidence of the downsides of conventional normalization in encompassing diversity, can be seen either

- as a transfer/generalization to all citation counts of Thompson-Dean ideas, implemented by Small & Sweeney in cocitation measure, firstly at the journal-level.

- or a restriction/ instantiation of the Pinski and Narin influence measures by dropping recursivity which causes an efficient but disputable over-weightin.

## 3. Productivity of distant linkages?

The high proportion of references internal to the discipline of bibliometrics/informetrics is not an artefact. Of course, this does not mean that the background landscape lacks diversity: a few steps backwards, in this particular article, as well as in many others, we would find uninterrupted flows from social sciences (diversity, reward system), physics-mathematics (network analysis), statistics (coining of indicators), etc. If I remember correctly we were also limited by the rules on bibliography length applicable for Jasist "notes" (this was not a full article) and had to sacrifice some weak-ties references. Anyway, the article cannot be credited for having established those distant linkages which shaped its background landscape. The long-distance communication wormholes were already opened and firmly established by Simon, Price, Garfield, Narin...

## 8.6    Appendix for Henry Small's email regarding co-citation analysis

Dear Lutz and Iman,

The references in my paper really don't tell the full story. In my case you could argue that I was combining ideas from two fields (history of science and information science) but this was not apparent from my references which are mostly in information science, with two from physics. But the physics ones reflect the data I was using much as a historian of science would cite scientific paper in telling the history of some topic and so are misleading. The co-citation idea really came out of work I was doing in the history of science in my previous job at AIP's Niels Bohr Library in New York where I was working on a project to document the history of nuclear of physics. I have written about this before. Anyway, my boss on this project, Charles Weiner, was a historian and gave me the job of figuring out what nuclear physics was back in the 1920s and '30s, who were the important people were and what they discovered. This was to serve as a guide to what they would need to document in their project [the Niels Bohr Library is an archive that collects papers of physicists, oral history interviews, etc.]. So what I did was started collecting bibliographic data on physics using Physics Abstracts (it was actually called Science Abstracts back then). For each papers I found under what I considered a "nuclear physics" headings I collected subject headings assigned, author and institution names, key words, references, etc. I was really doing bibliometrics without really knowing it. At some point I thought I might be able to pull all this information together in graphical form, like a network diagram of physics. I did subject heading networks, author networks, key word, and reference networks, etc. I would do what I called "pairings" of the difference data elements, subject headings jointly assigned to papers, key words jointly occurring, and what I called reference pairings. I wrote all this up in a report which was never published. I did try to publish one paper in a history of science journal but it was rejected! Of all the "pairings" I attempted I thought the subject heading and key words worked the best. The reference pairings didn't work very well probably because I had only used a small sample of references from each paper.

When the project terminated I had to find another job and I decided I wanted to work at place where they collected this kind of data. I wrote to practically every one of the A&I services at that time and ISI was the only organization that responded. When I started work at ISI in 1972 my first thought was to try to use key word pairs. However, I soon discovered that in a multidisciplinary database like the SCI there were massive problems with homography – the same words that mean different things in difference fields, like "cell" or "plasma". So I turned back to using references and began looking at the citation index, and was surprised to find strong patterns of what I then dubbed co-citation. That's how it happened.

You could argue that I was combining ideas from information science and history of science, but this would only be apparent if you knew about my work history and my career path. I was really very much an outsider coming to information science but I had been talking to information science people at the AIP because they had a physics information division. And my approach to history of science was not in the mainstream either and not welcomed by that field. In a sense I was an outsider in both fields. I don't know if this helps. Let me know if you need any further information.

Good luck,

Henry

### 8.7 Appendix for Jorge Hirsch's email regarding h-index

Dear Lutz and Iman,

My paper is probably not very representative in this respect. I was not working in the field of bibliometrics and was unfamiliar with the literature. I was always interested in citations as an objective way to evaluate the importance of scientific contributions, around 2003 I came to the idea of the h-index while evaluating applications for positions in my department as a simple way to summarize the information contained in citation records. I started using it and studying its properties but wasn't thinking about publishing a paper on it. I also shared it with some colleagues who found it interesting. Then in early 2005 I read the paper by Redner in Physics Today (ref. 3 in my paper), it stimulated my interest in bibliometrics and got me thinking that maybe I should write up something on the h-index. At that point and later I read some of the other references cited in my paper and other papers in the field of biblometrics but I don't think they played a major role in my thinking.

I hope these comments are helpful. Best regards,

Jorge

## 8.8 Appendix for Jason Priem's email regarding Altmetrics

Hi Lutz,

I'm on a plane and going to try to type up what I can remember. I have to reiterate that am really just so honored that you'd be interested at all. So, here we go. Apologies for informal grammar or spelling errors as I'm going to try to just type this all out real fast without worrying too much about editing.

Before I did anything with info science, I was a middle school teacher. This was the early 2000s, and there was a growing "edu-blogosphere" of teachers with blogs that I got really interested in, and ended up following this online community really closely.

For various reasons I decided I wanted to give teaching a break and go back to school. I wanted to do something with studying online communities, because I'd gotten so much from this online teacher blog community and was fascinated by it. But I had zero qualifications (my degrees were in history and education) so I figured I needed to do some kind of really cool technical project to convince a doctoral program to let me in.

So, I taught myself to code and made this thing called FeedVis, which was a visualization of the edu-blogosphere that tried to capture what they were talking about at any given moment, and track that how that conversation changed over time. You could put any list of RSS feeds in it as well, so you could use it for any online blog community. What I was really excited about was the idea of telling a community about itself, in ways that any individual member of the community wouldn't be able to see. Sort of a birds-eye view of the conversation.

I ended up at UNC SILS, and I learned about this thing called bibliometrics and got really interested in it. This was the same thing I had been trying to do with FeedVis, but instead of understanding the conversational topology of some blog network, it was trying to understand and map the frontiers of human knowledge. This, I thought, was way way more exciting.

So I dove into learning about bibliometrics. But as I learned more I felt that a lot of the most interesting work understanding the citation graph had already been done. Of course I'm sure many people in the field would disagree with me on this! It's certainly true that I lacked the technical knowledge to really appreciate the more nuanced work still being done from this angle.

But for whatever reason, I was not that excited about doing more work on citations, even though I was still really excited about scientometrics overall. I just felt that in terms of my interest, that particular vein of ore had been largely tapped out. So I went looking for other ways to map or analyze the scholarly conversation.

At first, I spent a lot of time looking into argument mapping, work that was trying to construct networks of claims and evidence rather than networks of papers. This was really cool because it's a more nuanced way of doing the same thing citation networks capture only very roughly. We could in theory build a network of claims and counter-claims that would look very much like the citation graph, and be closer to what's actually happening when we do scholarship (citing papers as a whole is very coarse).

But I began to feel that it was too hard to construct that network. It seemed to me that approaches relying on manual annotation/markup of arguments were to labor-intensive to work at scale, and automated approaches didn't yet seem mature enough (this latter point is changing I think as the machine learning renaissance blossoms, and I have high hopes for the future here).

So I was talking to my advisor, Brad Hemminger, about maybe using article readership to track impact, and he mentioned that PLOS was doing some interesting work sharing metrics like that.

This was PLOS's ALM (article-level metrics) program. As I looked into this I got pretty interested. This was what I had been looking for! This was a legitimately new, unexamined source of data, a new instrument with which to observe scholarly communication that had been invisible before. I think PLOS deserves enormous credit for leading the way on altmetrics.

So I started collecting some data using PLOS ALM dataset, and I got real excited about it, and started combing the lit to see who else was talking about this.

I read a lot of Blaise Cronin's work, and realized that he'd really seen all this coming and was way way ahead of me. And I loved his writing so much, that it gave me a lot of encouragement that this really smart guy thought there was a future in using the web to find "scientific street cred" (in his words). And I found Mike Thelwall's webometrics work, which was also encouraging. Both because he was using the web like I wanted to, and because it seemed clear to me that social media sources held more promise than the more "Web 1.0" sources he'd been using. Both Blaise and Mike were way ahead of me and I suspect would've done everything I did first (but better) if their careers had been timed a little differently. I'm glad I got a chance to do some of that, and hope I didn't let them down. And of course they are both continuing to put out great work, some of it in the area of altmetrics now (which feels me with much delight).

I also remember that along with doing a lot of reading, I bought lunch for Cassidy Sugimoto, who was then a SILS [School of Information and Library Science, University of North Carolina at Chapel Hill] doctoral student a few years ahead of me, in order to get her opinion about this stuff. So I kind of laid it all out for her, because I thought this could be big, but wanted to see if I was crazy. She was encouraging, which helped. She also ended up doing some altmetrics work later, which again makes me so happy. Here my memory is a bit fuzzy. I do recall that I hated the name "article level metrics" because it's talking about the level of aggregation rather than the type of indicator. Citations can be article-level, and bibliometrics has been looking at those kinds of "ALMs" forever. So I thought (and think) it was just a very bad name if we wanted to push these ideas forward. We needed a name that characterized the type of indicator. Which was going to be a challenge because they were and are very diverse.

So I tweeted about "altmetrics" to see if that name would get any traction. SILS definitely had a culture that expected grad students to be contributing to the field, early and often. I'm glad they did (and do). As I recall I thought of that name in the shower, and tweeted it right away afterwards because, why not.

I also figured there was no way the bibliometrics community would ever accept such a silly name so I came up with *another* name, and I published an article in First Monday about "scientometrics 2.0." Happily for everyone that did not catch on.

So I was starting to doing actual altmetrics research at that point, and the idea was beginning to catch on. I was really getting passionate about the idea at this point…not just altmetrics as a new data source for scientometrics, but as a part of a brand new way to organize scholarly communication, as a way to scale up the filtering function of peer review to be as big and fast as the web. And I felt it just needed a real strong publicity push so that other people would get excited about it and start doing the research and building the tools. So I had the idea right away that I should write a "manifesto." I minored in art history as an undergrad, and 20th century art movements were always doing these manifestoes and I'd thought then that that was so cool and rebellious and dramatic, so I'd always wanted to do one. So "aha I'll write a Manifesto" was something I needed very little excuse to think.

But I figured I'm just some PhD student no one's ever heard of, so no one's going to listen to me. I reckoned that if I had some coauthors that people respected, well maybe that's the kind of thing people would listen to. So I wrote the whole thing at ASIST [Association for Information Science and Technology] actually, in between sessions, running back and forth up to my hotel room to write…I was sharing that room with three other grad students and it was just chaos. Good times. And I invited three coauthors, none of whom were information scientists, but all of whom had written altmetrics papers of some kind Before It Was Cool. And they were kind enough to accept, and give some great thoughts on the manuscript and how best to present it.

And I think that's it! Apologies on the length. I set out to do as comprehensive a reminiscence as I could…since I was not at all on time, at least I could be thorough. And I'm sitting on a plane with no wifi so for once have plenty of time to write :)